\documentclass[proceedings, preprint]{rmaa}

\usepackage{paralist}

\usepackage{psfrag,color}

\SetYear{2008}
\SetConfTitle{Astronom\'ia Din\'amica en Latinoam\'erica}

\title{Environment and mass dependencies of galactic $\lambda$ spin
parameter: cosmological simulations and SDSS galaxies compared} 

\author{
  B. Cervantes-Sodi,\altaffilmark{1}
  X. Hernandez,\altaffilmark{1} 
  Changbom Park, \altaffilmark{2} \&
  Juhan Kim\altaffilmark{2}}

\altaffiltext{1}{Instituto de Astronom\'\i a,
Universidad Nacional Aut\'onoma de M\'exico
A. P. 70--264,  M\'exico 04510 D.F., M\'exico (xavier,bcervant@astroscu.unam.mx).}

\altaffiltext{2}{Korea Institute for Advanced Study, Dongdaemun-gu, Seoul 130-722, Korea.}

\shortauthor{B. Cervantes-Sodi et al.}
\shorttitle{Environmental and mass dependencies of $\lambda$}

\listofauthors{B. Cervantes-Sodi, X. Hernandez, Changbom Park, \& Juhan Kim}
\indexauthor{Cervantes-Sodi, B.}
\indexauthor{Hernandez, X.}
\indexauthor{Park, C.}
\indexauthor{Kim, J.}

\abstract{We use a sample of galaxies from the Sloan Digital Sky Survey (SDSS) to search for correlations between the $\lambda$ 
spin parameter and the environment and mass of galaxies. In order to calculate the total value of $\lambda$ 
for each observed galaxy, we employed a simple model of the dynamical structure of the galaxies which allows
a rough estimate of the value of $\lambda$ using only readily obtainable observables from the luminous galaxies. 
Use of a large volume limited sample (upwards of 11,000) allows reliable inferences of mean values and dispersions 
of $\lambda$ distributions. We find, in agreement with some
N-body cosmological simulations, no significant dependence of $\lambda$ on the environmental density of the galaxies.
For the case of mass, our results show a marked correlation with $\lambda$, in the sense that low mass
galaxies present both higher mean values of $\lambda$ and associated dispersions, than high mass galaxies. This last direct empirical result, at odds with expectations from N-body cosmological simulations, provides interesting 
constrain on the mechanisms of galaxy formation and acquisition of angular momentum, a valuable test for cosmological models.
}

\addkeyword{galaxies:formation}
\addkeyword{galaxies: statistics}
\addkeyword{galaxies: structure}
\addkeyword{cosmology: observations}

\begin{document}
\maketitle

\section{Introduction}
\label{sec:intro}

One of the most studied parameters in numerical simulations of
formation and evolution of galaxies, is the $\lambda$ spin parameter,
first introduced by Peebles in 1969, where it was used to test the
gravitational instability picture, as a theory for the origin of
galactic angular momentum. Since then, there have been several studies using
the $\lambda$ parameter as an indispensable tool of analysis, or characterizing its
distribution in cosmological N-body simulations. In general, the distribution of this parameter,
coming from numerical simulations, is well fitted by a log-normal distribution, characterized
by two parameters; $\lambda_{0}$, the most probable value,
and $\sigma_{\lambda}$, which accounts for the spread of the
distribution. In a recent work, Shaw et al. (2006)
showed a compilation of results for estimates of these parameters, arising from numerical
studies performed by several authors, lying in the ranges: $0.03<\lambda_{0}<0.05$
and $0.48<\sigma_{\lambda}<0.66$, and some of us in a previous work
(Hernandez et al. 2007), using the same 
sample of 11 597 galaxies from the Sloan Digital Sky Survey (SDSS) we treat here, obtained
the distribution of this parameter, with results well fitted by a log-normal function with
parameters $\lambda_{0}=0.04\pm 0.005$ and $\sigma_{\lambda}=0.51 \pm 0.05$, for the first time
derived from a statistical sample of real galaxies.

With the recent explosion of high resolution cosmological
simulations, studies concerning the spin parameter begun to explore dependencies of the $\lambda$
parameter with other physical parameters, mainly the total mass of the galaxy and the environment in
which the galaxy is immersed. The results are as varied as the techniques employed to perform and analyse the
simulations, but  there is good agreement that for the case of a $\lambda$-environment relation,
it should be very weak or inexistent (Bett et al. 2007, Maccio et al 2007). The case of the
$\lambda$-mass relation presents more discrepancies between results coming from different groups, some
claiming the existence of a weak relation (Jang-Codell \& Hernquist 2001, Bett et al. 2007) and others not finding any (Maccio
et al. 2007). The aim of the present work is to derive distributions of the spin parameter as functions of mass
and environment, using a first order estimate of galactic halo $\lambda$ applied to a large sample of galaxies from
the Sloan Digital Sky Survey (SDSS), in order to constrain and infrom theories of angular momentum acquisition
and galaxy formation.

\section{Theoretical framework}
\label{sec:intro}

The angular momentum is commonly characterized by the dimensionless spin parameter: 

\begin{equation}
\label{Lamdef}
\lambda = \frac{L \mid E \mid^{1/2}}{G M^{5/2}}
\end{equation}

where $E$, $M$ and $L$ are the total energy, mass and angular momentum of the configuration, respectively. In
Hernandez \& Cervantes-Sodi (2006), we derived a simple estimate of $\lambda$ for disk galaxies, using a simple
model employing two components to describe the galaxy, a disk for the barionic component with an exponential surface mass density $\Sigma(r)$;

\begin{equation}
\label{Expprof}
\Sigma(r)=\Sigma_{0} e^{-r/R_{d}},
\end{equation} 

where $r$ is a radial coordinate and $\Sigma_{0}$ and $R_{d}$ are two constants which are allowed to vary from
galaxy to galaxy, and a dark matter halo having an isothermal density profile $\rho(r)$, responsible for establishing
a rigorously flat rotation curve $V_{d}$ throughout the entire galaxy;

\begin{equation}
\label{RhoHalo}
\rho(r)={{1}\over{4 \pi G}}  \left( {{V_{d}}\over{r}} \right)^{2}. 
\end{equation}

We assume: (1) that the specific angular momentum of the disc and halo are equal e.g. Fall \& Efstathiou (1980); 
(2) the total energy is dominated by that of the halo which is a virialized gravitational structure; (3) the disc 
mass is a constant fraction of the halo mass $F=M_{d}/M_{H}$. Introducing a Tully-Fisher relation and fixing constants with Galactic
parameters, we obtain:

\begin{equation}
\label{LamObs}
\lambda=21.8 \frac{R_{d}/kpc}{(V_{d}/km s^{-1})^{3/2}}.
\end{equation}

 The above equation allows a direct estimate of $\lambda$ for any disk galaxy. The corresponding equation for elliptical
galaxies was derived in Hernandez et al. (2007) modeling two components; a barionic matter distribution with a Hernquist density
profile and a dark matter halo showing no difference from that of disk galaxies (Treu et al. 2006). Considering the structure of the galaxy as
virialized, the mass is obtained in terms of the half light radius and the velocity dispresion $\sigma$. By dimensional analysis,
the specific angular momentum is characterized by $(GM_{b}a)^{1/2}$,
where $M_{b}$ is the baryonic mass and $a$ is the mayor axis. Introducing a numerical factor to account for the
dissipation of the baryonic component, its dynamical pressure support and the projection effects, we finally obtain

\begin{equation}
\label{LamObsEll}
\lambda= \frac{0.1173 e (a/kpc)^{2/7}}{(\sigma/km s^{-1})^{3/7}}.
\end{equation}

The accuracy of equation (4), was tested in Cervantes-Sodi et al. (2008), making use of numerical simulations, showing an
agreement between the estimate and the real value of $\lambda$ always better than 30 per cent, in most cases much better.

\section{Comparisons of $\lambda$ distributions from the SDSS and cosmological N-body simulations}

\subsection{Observational sample}

The sample of real galaxies employed for the study comes from the SDSS Data Release 5 (Adelman-McCarthy et al., 2007). It is a volume limited sample having galaxies in the redshifts interval $0.025 < z < 0.055$ and absolute magnitudes $M_{r}-5 log h \leq -18.5$. Since most of the studies concerning spin distributions from simulations presents their 
results at $z=0$, we limited the sample to low redshifts. This sample contains 32 550 galaxies
for which Choi et al. (2006) have determined the exponential disc scales, absolute magnitudes, velocity
dispersions, de Vacouleurs radii and seeing corrected isophotal ellipticities for each galaxy, assuming a $\Lambda CDM$ universe with 
$\Omega_{M}=0.27$, $\Omega_{\Lambda}=0.73$ and $h=0.71$. To obtain $\lambda$ for each galaxy, we need to discriminate
between elliptical and disc galaxies; in order to do that we used the prescription of Park \& Choi (2005), in 
which early (ellipticals and lenticulars) and late (spirals and irregulars) types are segregated in a $u - r$ colour versus $g - i$ 
colour gradient space and in the concentration index space. In order to apply equation ~\ref{LamObs} to the disc 
galaxies of the sample, we need the rotation velocity, which is inferred from the absolute magnitude using a TF 
relation, then, to avoid the problem of internal absorption in edge-on galaxies, we employed only spiral galaxies having axis ratios $> 0.6$ and inferred rotation velocities in the range of $80<V_{R}<430$, well within the range of applicability of the TF realtion we are using. After applying these two cuts, and removing a randomly selected fraction of ellipticals, so as to maintain the original early to late type fraction, we are left with a total of 11 597 galaxies.

For the elliptical galaxies we apply equation ~\ref{LamObsEll}, as fully described in Hernandez et al. (2007), using the corrected isophotal ellipticities to calculate the eccentricity.

\begin{figure*}[!t]

\begin{tabular}{cc}
\includegraphics[width=0.475\textwidth]{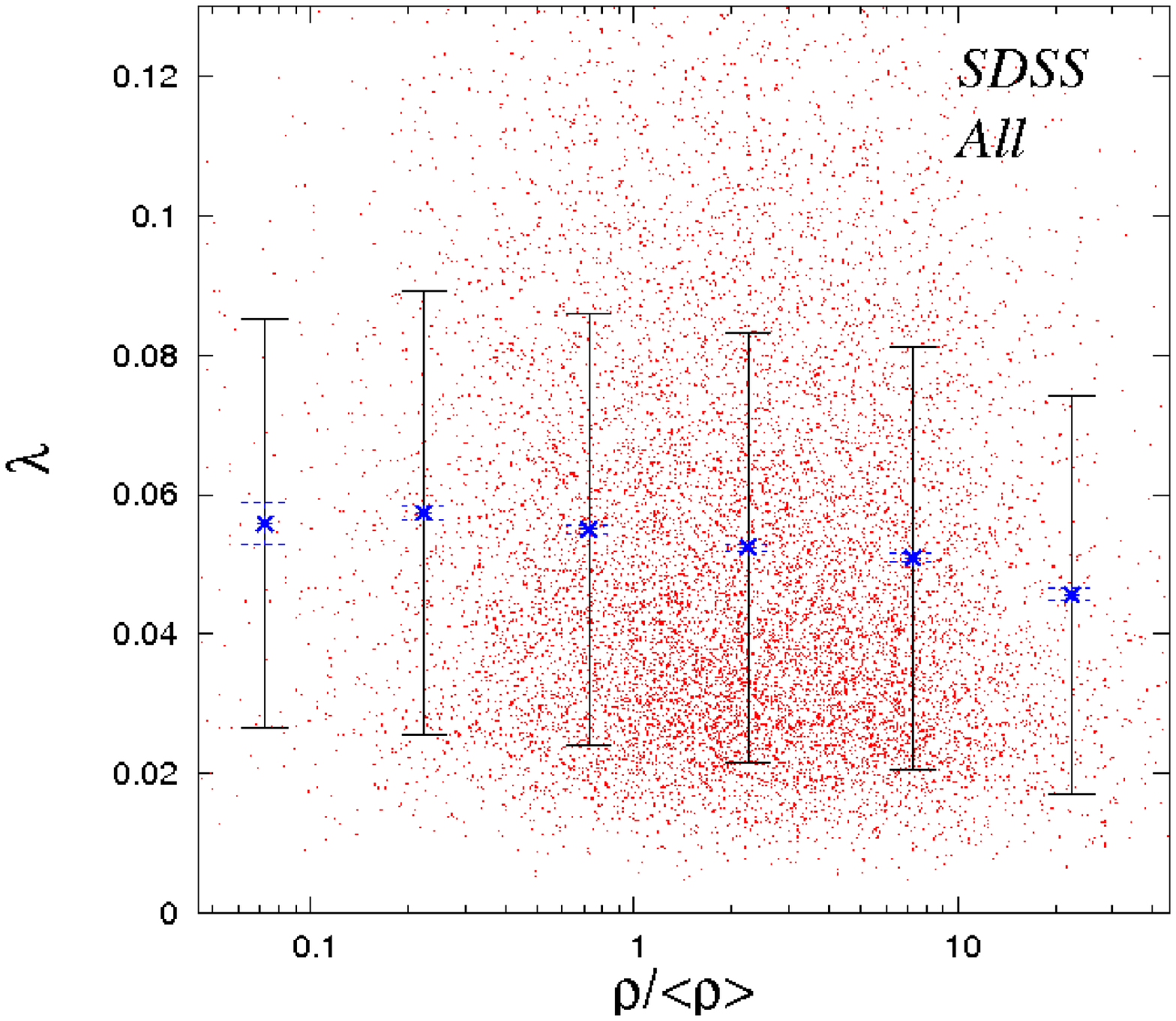} & \includegraphics[width=0.475\textwidth]{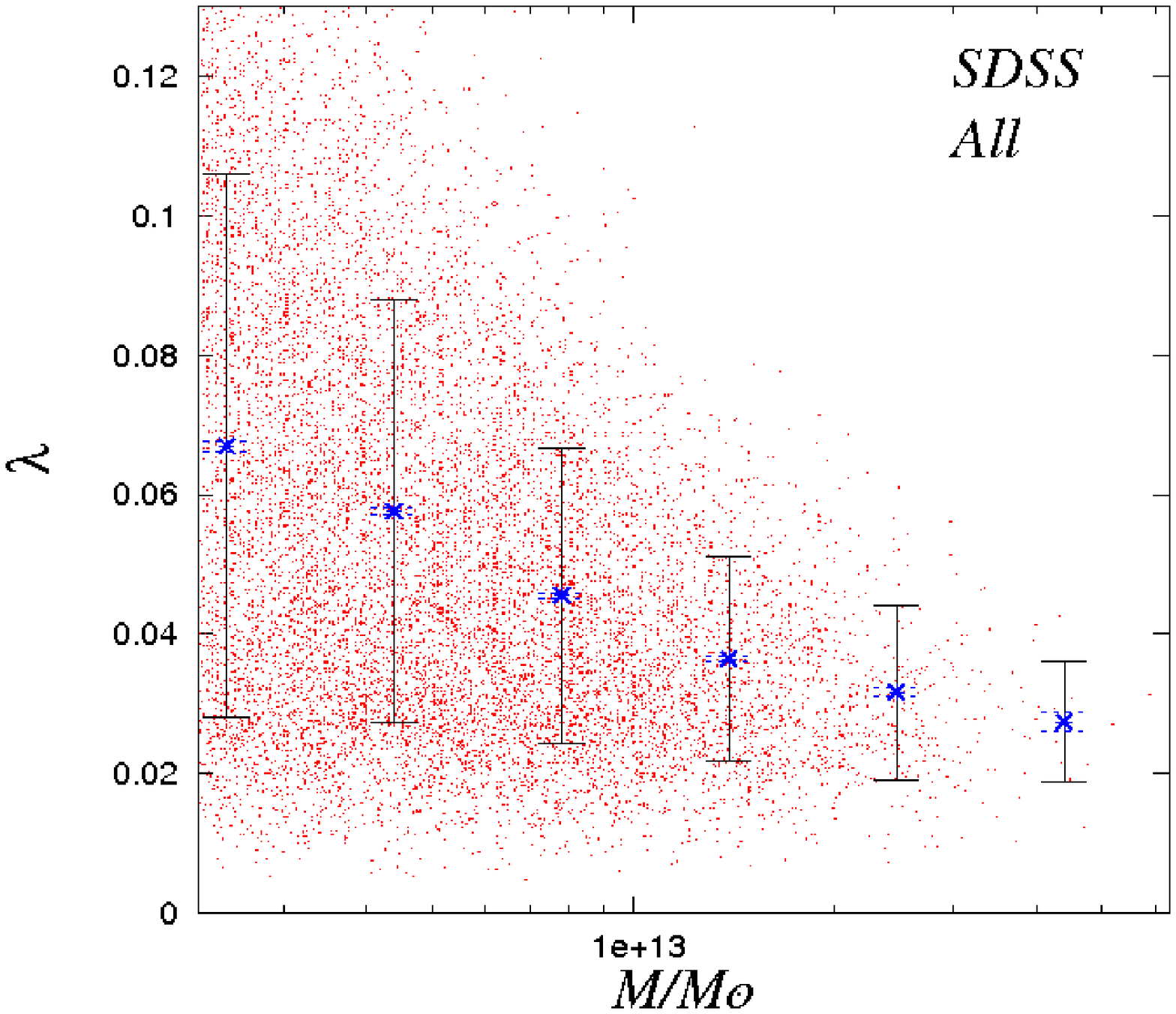} \\
\includegraphics[width=0.475\textwidth]{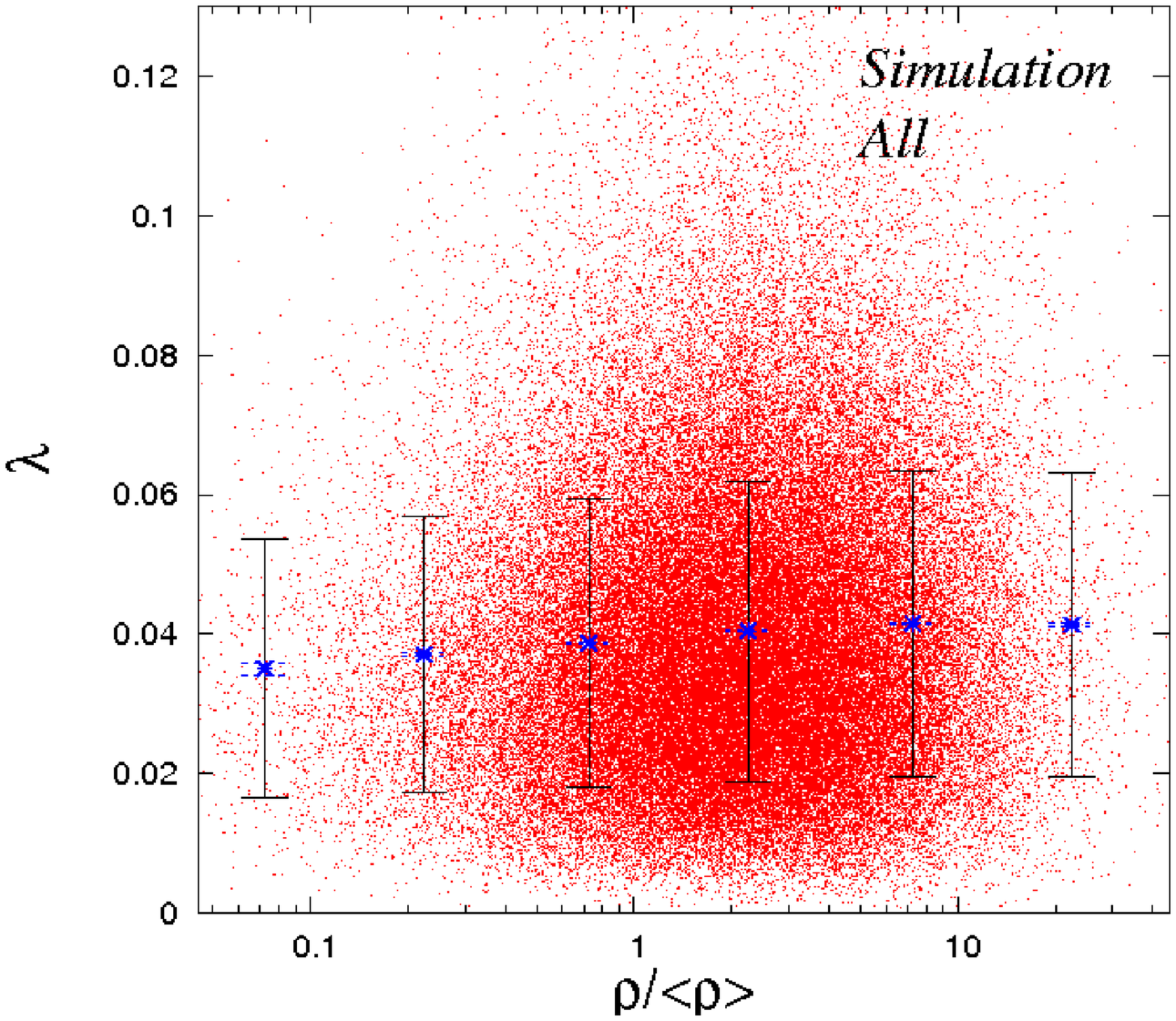} & \includegraphics[width=0.475\textwidth]{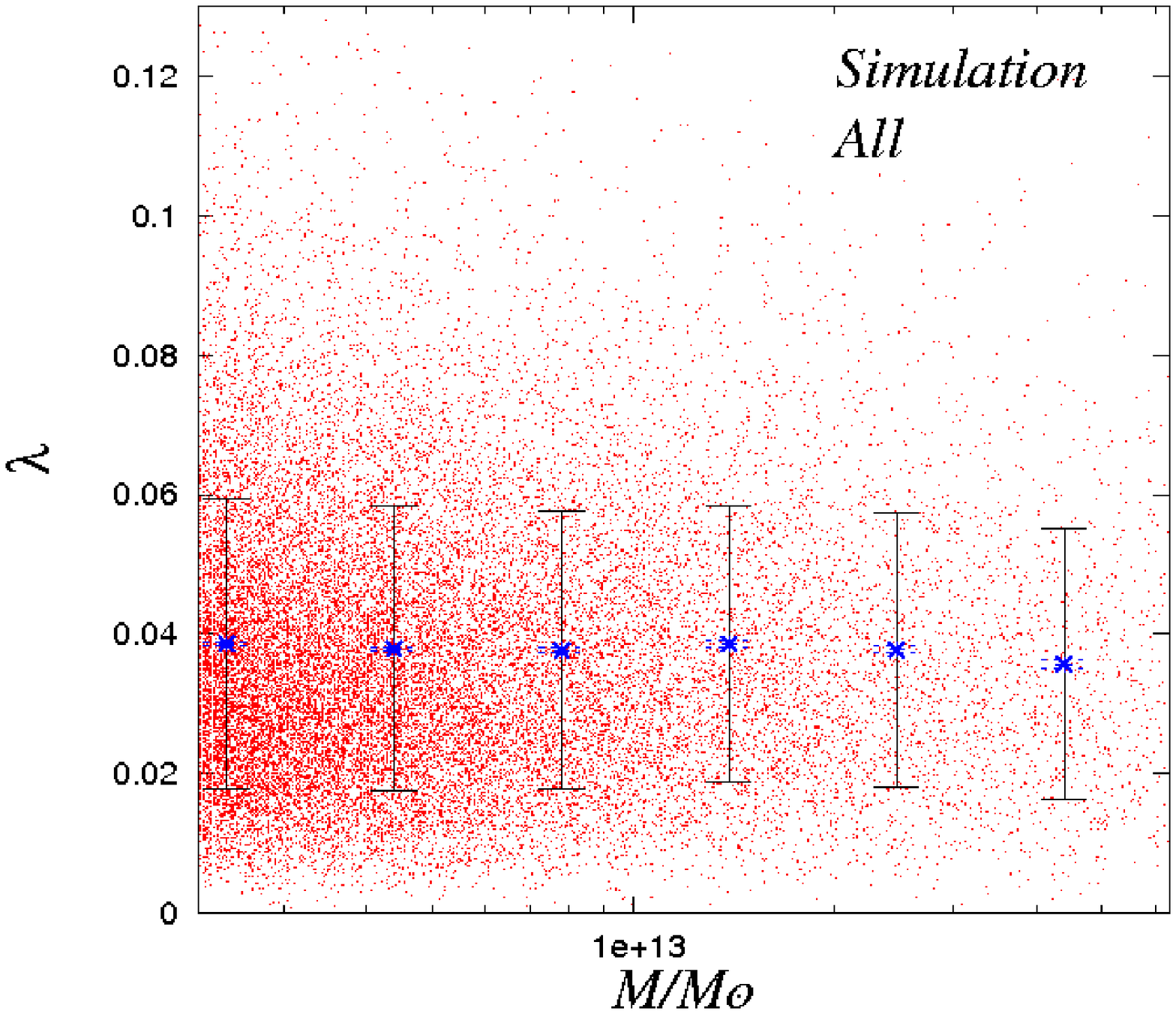}
\end{tabular}
\label{Density}
\caption[ ]{Relations between galactic spin and environmental density and mass, showing the mean $\lambda$ value for each bin as a function of 
the normalized density, the solid error bars correspond to the dispersion and the broken error bars to the standard error in the calculation of the mean value. \textit{Top} panels corresponding to the SDSS sample using the 11 597 galaxies, 
on the left-hand $\lambda$ as a function of environmental density, on the right-hand $\lambda$ as a function of mass. 
\textit{Bottom} panels corresponding to the sample of 100 000 simulated dark matter haloes, on the left-hand $\lambda$ as a function of environmental density and on the 
right-hand as a function of mass. }
\end{figure*} 

\subsection{Simulation  \& Subhalo Finding}
To compare our observational sample with a numerical simulation,
we have made a cosmological N-body simulation of a $\Lambda$CDM model, Dutton, A. A., van den Bosch, F. C., Morre, B., Potter, D., \& Stadel, J
in a cubic box with the side length of $614 h^{-1} {\rm Mpc}$. The model
parameters are $h=0.7$, $\Omega_m=0.27$, $\Omega_b=0.046$, and
$\Omega_\Lambda=0.73$, given by the WMAP 1-year data (Spergel et al. 2003).
The simulation gravitationally evolved $1024^3$ CDM particles from redshift
$z=48$ to $z=0$ taking 1880 global time steps.
The linear density fluctuations of matter at the present epoch is normalized
by  $\sigma_8 = 0.9$, the RMS density fluctuation at $8h^{-1}$Mpc for a tophat
window function.
The mass of each simulation particle is $M_p = 1.4 \times 10^{10} h^{-1}{\rm M_\odot}$
and the force spatial resolution is about $60 h^{-1}kpc$.

To identify subhalos in the simulation,
we first extracted the dark matter particles located in virialized regions
by applying the standard Friend-of-Friend (FoF) method
with a linking length equal to one fifth the mean particle separation. This
length is a characteristic scale for the identification of virialized
structures. To each FoF particle group, we applied the PSB method
(Kim \& Park 2006) to find subhalos.

For comparisons with
our observational results we randomly selected 100 000 halos from the simulation, for which the spin parameter $\lambda$ was calculated numerically. The normalized local
density parameter is obtained following the same method as used for our
SDSS sample (for more details
see Park et al. 2007).

\section{Results}
To study if there is any correlation between spin and the environment, we use an estimation of the local background density defined 
directly from observations, that is continuous and able to characterize the full range of galaxy environments. It measures 
the local number density of galaxies at a given smoothing scale (Park et al. 2007).  We divided both of our samples, the SDSS and the simulation, into 6 cuts according to the normalized environmental density $\rho/<\rho>$, and for each cut 
we obtained the mean $\lambda$ value and its dispersion. The result is presented in figure 1 left panels, where the value 
of total infered $\lambda$ versus $\rho/<\rho>$ for each galaxy is plotted and we show the mean $\lambda$ value for each cut as a function of 
the normalized density, the solid error bars corresponds to the dispersion and the broken error bars to the standard error in the calculation of the mean value, this convention will be kept in the following figure.
As can be seen, there is no clear trend, taking into account the dispersion in each bin, in neither the SDSS nor the simulation sample.

On the right panels of figure 1 are presented the plots of $\lambda$ versus $mass$ for the SDSS sample (top panel), and the simulation sample (bottom panel). We notice that for the SDSS sample, the mean value 
of $\lambda$ tends to decrease as the mass increases. But not only the mean value, also the dispersion of $\lambda$ tends to
 decrease as the value of the mass increases. The effect is stronger using only the disc galaxies, where our estimate 
is more precise (see Cervantes-Sodi et al. 2008). Large galaxies are seen to form a more coherent low $\lambda$ sample with small dispersion, while
small galaxies show mean values of $\lambda$ of about a factor of 3 larger than what is found for our most massive galaxies. 
The dispersion about mean values shows the same trend, with small galaxies in our SDSS sample having a much larger
dispersion than the large galaxies. The trend is not reproduced by
the simulation used in this work, and neither has it ever been reported, to the large extent seen in the SDSS sample, in any cosmological N-body simulation found in the literature. However, precisely such a trend was recently
confirmed by Berta et al. (2008) using a sample of 52 000 galaxies from the SDSS employing a similar method as the described
here to estimate $\lambda$. We take our results as evidence of a mechanism of acquisition
of angular momentum for galaxies, where it is the ambient tidal field what torques up a halo, with little participation
of repeated mergers. An extensive development of these ideas can be found in Cervantes-Sodi et al. (2008).


\begin{thebibliography}

\bibitem{}Adelman-McCarthy, J. K., et al.\ 2007, \apjs, 172, 634
\bibitem{}Bett, P., Eke, V., Frenk, C. S., Jenkins, A., Helly J., \& Navarro, J.\ 2007, \mnras, 376, 215
\bibitem{}Berta, Z., Jimenez, R., Heavens, A. F., \& Panter, B.\ 2008, \mnras, submitted (astro-ph/0802.1934)
\bibitem{}Cervantes-Sodi, B., Hernandez, X., Park, C., \& Kim, J.\ 2008, \mnras, accepted (astro-ph/0712.0842)
\bibitem{}Choi, Y., Park, C., \& Vogeley, M. S.\ 2007, \apj, 658, 884
\bibitem{}Fall, S. M., \& Efstathiou, G.\ 1980, \mnras, 193, 189
\bibitem{}Hernandez, X., \& Cervantes-Sodi, B.\ 2006, \mnras, 368, 351
\bibitem{}Hernandez, X., Park, C., Cervantes-Sodi, B., \& Choi Y.\ 2007, \mnras, 375, 163
\bibitem{}Jang-Codell, H., \& Hernquist, L.\ 2001, \apj, 548, 68
\bibitem{}Kim, J., \& Park, C.\ 2006, \apj, 639, 600
\bibitem{}Maccio, A. V. et al.\ 2007, \mnras, 378, 55
\bibitem{}Park, C., \& Choi, Y.\ 2005, \apj, 635, L29
\bibitem{}Park, C., Choi, Y., Vogeley, M. S., Gott, J. R., \& Blanton, M. R.\ 2007, \apj, 658, 898
\bibitem{}Peebles, P. J. E.\ 1969, \apj, 155, 393
\bibitem{}Shaw, L. D., Weller, J., Ostriker, J. P., \& Bode, P.\ 2006, \apj, 646, 815
\bibitem{}Treu, T., Koopmans, L. V., Bolton, A. S., Burles, S., \& Moustakas, L. A. \ 2006, \apj, 640, 662 

\end{thebibliography}
\end{document}